\documentclass[12pt]{article}
\usepackage{epsfig}
\usepackage{graphicx}
\begin{document}
\begin{sloppypar}
\begin{center}
{\Large\bf New Theoretical and Experimental Correlation Aspects of
QCD-Instantons in High Energy Collisions} \vspace{0.5cm}

{\large V.I.Kuvshinov$^1$, V.I.Kashkan$^2$, R.G.Shulyakovsky$^3$}
\vspace{0.3cm}

{\it Institute of Physics, National Academy of Sciences of Belarus \\
F.Skorina av., 68, 220072 Minsk, BELARUS} \\
$^1$kashkan@dragon.bas-net.by \\
$^2$kuvshino@dragon.bas-net.by \\
$^3$shul@dragon.bas-net.by

\vspace{0.4cm} {\bf Abstract}
\end{center}
\noindent Possibility of experimental identification of
QCD-instantons in high energy collisions is studied by means of
correlations analysis in final states. Instanton-induced processes
amplitudes are performed in the framework of QCD in Gauss
approximation. Hadronization is taken into account by Monte-Carlo
method. Obtained results can be used as additional criterions of
QCD-instanton identification at HERA (DESY). Unlike previous
results in this report we consider also nonzero quarks modes
contribution into parton distributions of instanton processes.

\section{Introduction}
A possibility of strong growth of the cross-section of the {\it
instanton} transitions in high energy collisions was mentioned
first for \underline{electroweak} theory~\footnote{in electroweak
theory these processes can become observable at energy about
$O(10)$ TeV}~\cite{Ring90}. Shortly after this it was
shown~\cite{Bal} that \underline{QCD-instantons} can appear as a
new channel of deep inelastic scattering (DIS) and be (in
principle) observed at the present-day experiments unlike
electroweak instantons. Specifically, QCD-instantons can be
produced in quark-gluon subprocess  at HERA (DESY) (Fig.1).
\hspace*{3cm}
\begin{center}
\epsfxsize=2.5in \epsfysize=2.5in
\epsfbox{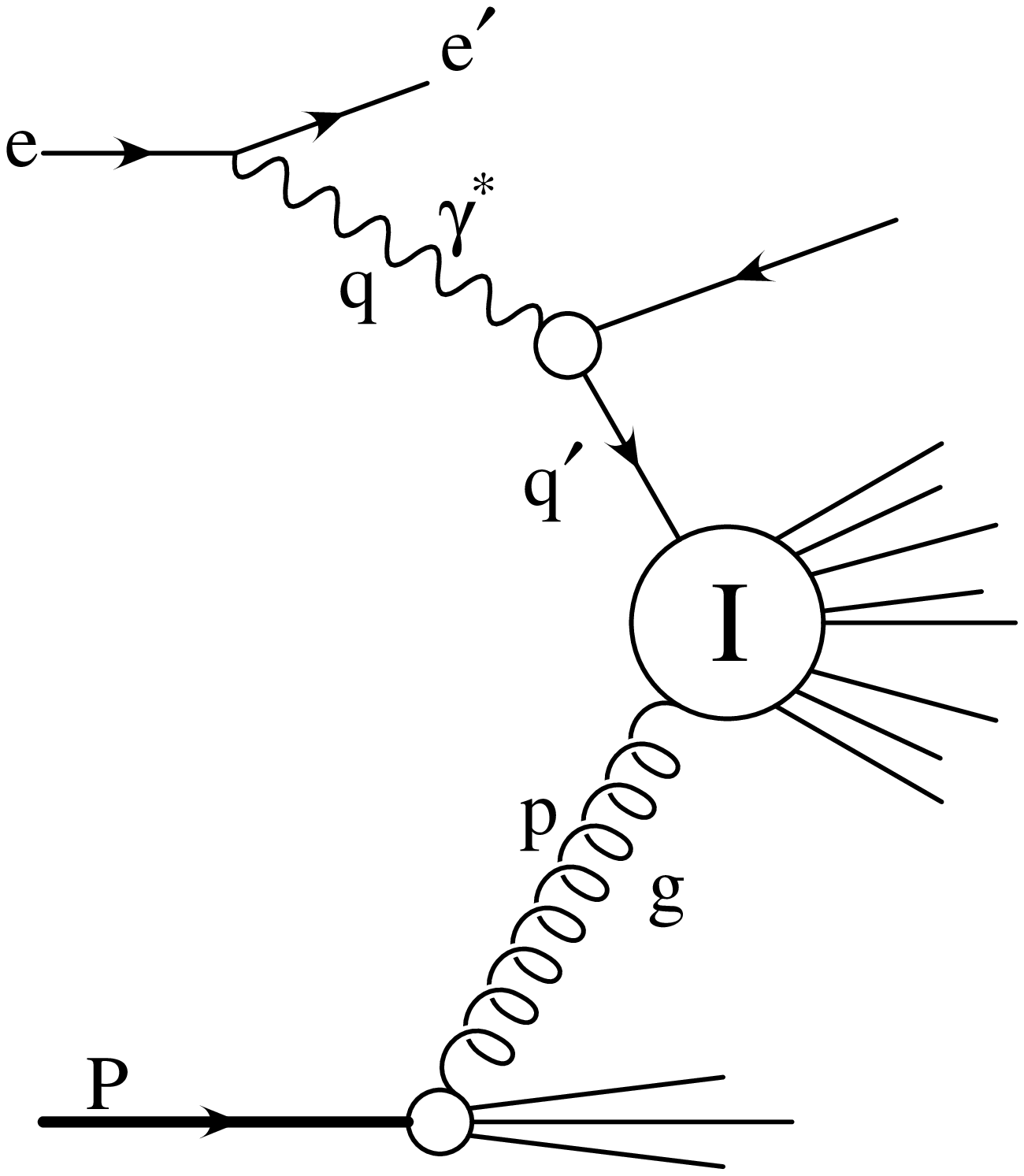}\end{center}
\vspace*{-12.6cm}

\vspace*{13cm}
\begin{center}
Fig.1. Instanton induced DIS at HERA.\end{center}
 \vspace*{0.1cm}

A set of important features of the process (large number of
secondary particles, specific behavior of cross-section and
structure functions, large transversal energy flow and others) was
already discussed by Schrempp, Ringwald and
collaborators~\cite{RSh,QCDINS}. It was shown that instanton
fraction in DIS can reach 1\%, but uncertainty inherent
calculations in QCD does not allow to consider confidently that
instanton channel in DIS is discovered.

\section{Correlation criterions of QCD-instantons in DIS}
We consider that analysis of correlations in DIS may be helpful
for the solving of the problem of experimental identification of
QCD-instanton. Preliminary results (calculation of 2-particle
correlation function~\cite{Acta1}, factorial and
$H_q$-moments~\cite{Acta2}) showed that instanton-induced
processes are characterized by specific form of correlation
characteristics at parton level. Footprints of these features
persist after hadronization. In particular normalized factorial
moments for instanton processes grow very slowly (Fig.2),
$H_q$-moments are characterized by first minima at $q=2$ unlike
ordinary DIS (Fig.3)~\cite{KSh}. Hadronization was taken into
account by means of Monte-Carlo package QCDINS~\cite{QCDINS} (the
programm which generates QCD-instanton-induced events).

\vspace*{0.6cm} 
\begin{minipage}{7cm}
\begin{center}
\epsfxsize=2.5in \epsfysize=2.5in \epsfbox{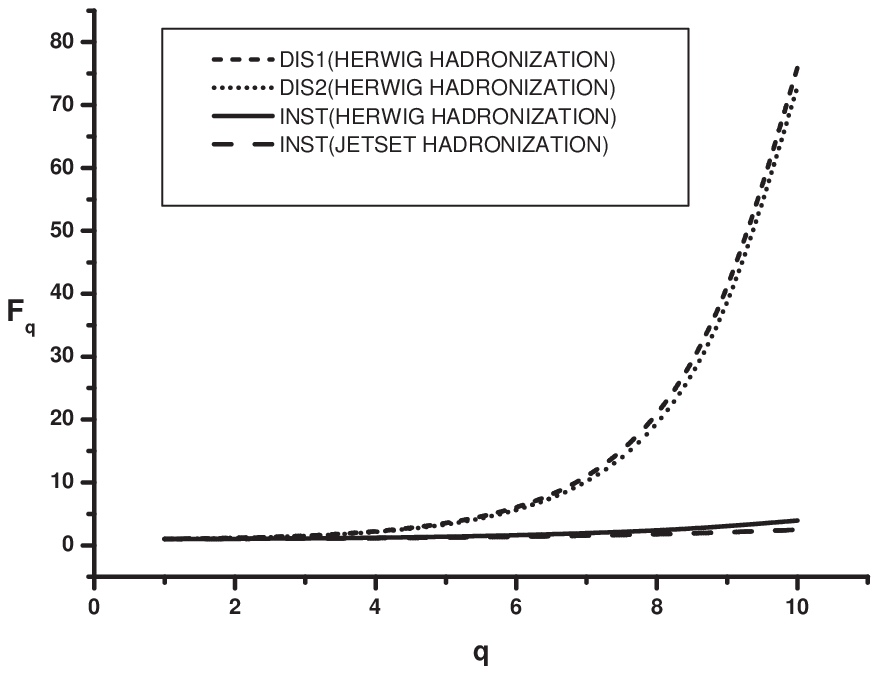}\end{center}
\end{minipage}

\vspace*{-6.5cm} \hspace*{8cm}
\begin{minipage}{7cm}
\begin{center}
\epsfxsize=2.5in \epsfysize=2.5in \epsfbox{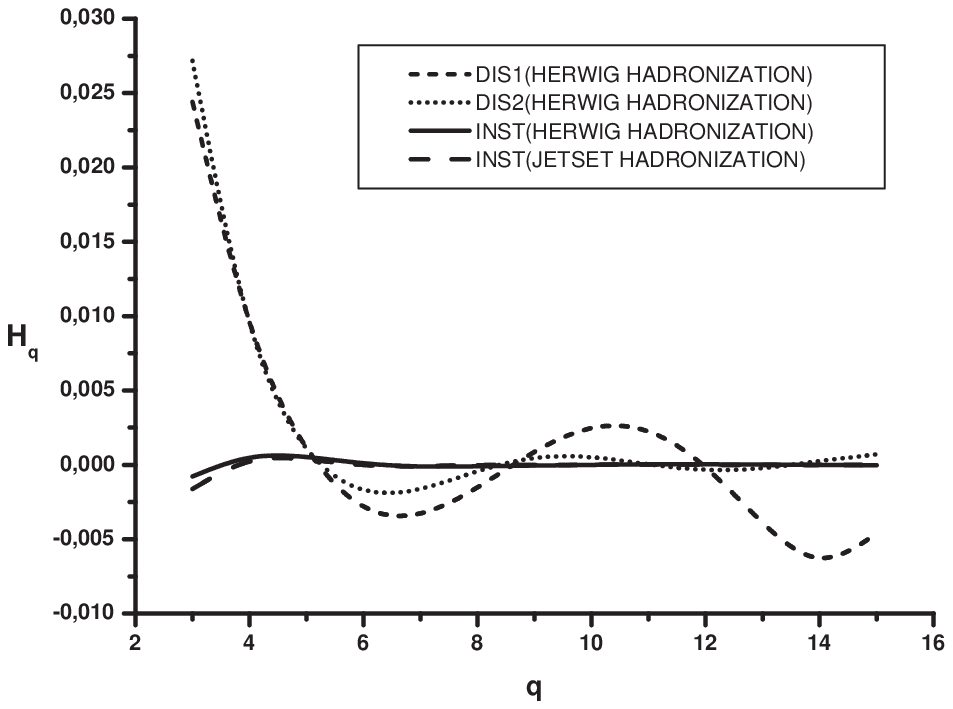}\end{center}
\end{minipage}

\vspace*{0.2cm}
\begin{minipage}{7cm}
\begin{center}
Fig.2. $F_q$ for one-jet (DIS1), 2-jet (DIS2) and instanton
induced DIS (INST).
\end{center}
\end{minipage}
\vspace*{0.1cm}

\vspace*{-1.6cm} \hspace{8cm}
\begin{minipage}{7cm}
\begin{center}
Fig.3. $H_q$ for one-jet (DIS1), 2-jet (DIS2) and instanton
induced DIS (INST).
\end{center}
\end{minipage}
\vspace*{0.1cm}

Thus correlation properties of instanton-induced DIS can be
considered as new criterions of the QCD-instanton identification
in addition to criterions of Schrempp, Ringwald et
al.~\cite{RSh,QCDINS}.

\section{Contribution of nonzero quark modes}
Usually it is supposed that only minimal number of quarks is
produced after "decay" of instanton (number of final gluons is
supposed to be arbitrary):
\begin{equation}\label{process}
q+g\to (2n_f-1)q+n_gg.
\end{equation}
This supposition was used by authors of the package
QCDINS~\cite{RSh,QCDINS} as well as in Ref.~\cite{KSh}.

Let us we consider instanton-induced processes with arbitrary
number of quarks. Contribution of these processes is determined by
nonzero fermion propagator~\cite{Shuryak}
$$
S^{nz}(x,y)=
\frac{1}{\sqrt{1+\rho^2/x^2}}\frac{1}{\sqrt{1+\rho^2/y^2}}\Biggl[\frac{(x-y)_{\mu}\sigma_{\mu}}{2\pi^2(x-y)^4}
\biggl(1+\rho^2\frac{x_{\nu}\sigma_{\nu}y_{\kappa}\bar{\sigma}_{\kappa}}{x^2y^2}
\biggr)-
$$
\begin{equation}\label{nonzero}-\frac{\rho^2}{4\pi^2(x-y)^2x^2y^2}
\biggl(\bar{\sigma}_{\mu}\frac{x_{\nu}\sigma_{\nu}\bar{\sigma}_{\mu}(x-y)_{\lambda}\sigma_{\lambda}y_{\omega}
\bar{\sigma}_{\omega}}{\rho^2+x^2}+\sigma_{\mu}
\frac{x_{\nu}\sigma_{\nu}(x-y)_{\lambda}\bar{\sigma}_{\lambda}\sigma_{\mu}y_{\omega}
\bar{\sigma}_{\omega}}{\rho^2+y^2}\biggr)\Biggr],
\end{equation}
where $\rho$ is instanton size, $\sigma_{\mu}=(-i\sigma_a,I),\,
\bar{\sigma}_{\mu}=(i\sigma_a,I)$.

After calculation\footnote{for detailed explanation of calculation
of the distribution of multiplicity on quarks produced after
instanton "decay" see~\cite{me}} we obtain Poisson distribution on
number of quark pairs (for every light quarks), which are produced
in the instanton processes:
\begin{equation}\label{Dist}
P_n=[e^{\xi^2}-1]^{-1}\frac{\xi^{2n}}{n!},\quad
\xi\approx\frac{1-x'}{x'},
\end{equation}
where Bjorken variable of instanton subprocess $x'>0.5$. Average
number of quark pairs for small $\xi$ reads
\begin{equation}\label{Dist}
<n>\approx 3(1+\xi^2).
\end{equation}
Contribution of non-zero modes can lead to another behavior of
characteristics of instanton processes and be important for the
experimental search of QCD-instantons. Monte-Carlo simulation is
in progress.

\end{sloppypar}
\end{document}